\documentclass[12pt]{JHEP}
\usepackage{amssymb,amsfonts}
\usepackage{amsmath}
\usepackage{fancybox}
\usepackage{frame}
\def\hri#1#2{\href{http://arxiv.org/abs/#1}{[ArXiv:#1]#2}}
\def\hre#1#2{\href{http://arxiv.org/abs/#1/#2}{[ArXiv:#1/#2]}}

\def\pa{\partial}

\def\nn{\nonumber}

\relax
\renewcommand{\theequation}{\arabic{section}.\arabic{equation}}
\def\be{\begin{equation}}
\def\ee{\end{equation}}
\def\bea{\begin{eqnarray}}
\def\eea{\end{eqnarray}}

\newcommand{\bear}{\begin{eqnarray}}

\newcommand{\eear}{\end{eqnarray}}
\newbox\pippobox

\def\b{\beta}
\def\g{\gamma}
\def\d{\delta}
\def\l{\lambda}

\def\p{\partial}
\def\nn{\nonumber}

\def\half{\frac12}

\def\co{{\cal O}}

\def\m{\mu}
\def\n{\nu}
\def\r{\rho}
\def\s{\sigma}

\def\sp{\;\;\;,\;\;\;}

\def\sq
\def\y{\psi}

\title{Holographic RG Flows and Nearly-Marginal Operators}

\author{\Large Jun Bourdier$^a$ and Elias Kiritsis$^{b,c,d}$\\
~\\
~\\
$^a$Department of Mathematics, King's College London, The Strand, WC2R 2LS, London, United-Kingdom\\
~\\
$^b$ \href{http://www.apc.univ-paris7.fr}{APC, Universit\'e Paris 7, Diderot},
 CNRS/IN2P3, CEA/IRFU, Obs. de Paris, Sorbonne Paris Cit\'e,
  B\^atiment Condorcet, F-75205, Paris Cedex 13, France (UMR du CNRS 7164)\\
  ~\\
$^c$\href{http://wwwth.cern.ch/}{Theory Group, Physics Department, CERN}, CH-1211, Geneva 23, Switzerland\\
~\\
$^d$ \href{http://hep.physics.uoc.gr}{Crete Center for Theoretical Physics},
Department of Physics, University of Crete, 71003 Heraklion, Greece
}

\preprint{CCTP-2013-16 \\ CCQCN-2013-11\\ CERN-PH-TH/2013-265 \\ KCL-MTH-13-10}      

\abstract{The holographic renormalization group flows associated with marginally relevant operators are analysed. The associated perturbative and non-perturbative $\beta$-functions are calculated and the consistent scalar potentials are identified. The presence of a Zamolodchikov metric in the multiscalar case is shown to affect $\beta$-functions starting at two loops.    }


\begin{document}

\def\g{\gamma}
\def\go{\g_{00}}
\def\gi{\g_{ii}}

\maketitle 

\section{Introduction and Results}

The holographic correspondence has revolutionized the way we think about both quantum field theory and string theory. Moreover, it has indicated that the two are intimately related, and that string theory can be viewed as the dynamics of sources of quantum field theory (QFT), \cite{LV,string,sslee}, namely  a form of quantum renormalization group.

A particular item of the QFT tool box, renormalization group (RG) flows have an elegant description in the string theory/gravity language. They correspond to bulk solutions of the equations of motion with appropriate boundary conditions. There is a way of writing the bulk equations using the Hamilton-Jacobi formalism, so that they look formally similar to RG equations, \cite{deboer,papa,bianchi}, a fact that has been exploited in numerous situations. There have been various takes on the form of the RG/coarse graining procedure in the holographic case, \cite{holorg}, including the Wilsonian approach to IR physics \cite{cgkkm}.

It is not yet known how to generalize the Hamilton-Jacobi formalism to cases where we are discussing non-trivial states in QFT, namely states with non-trivial temperature and finite density.
However, even in that case there are other techniques that allow the calculation of effective potentials, \cite{eff}, and effective actions, \cite{nitti}, along with  the related RG equations.

An important set of flows in QFT are generated by nearly marginal (marginally relevant) operators. This is the example of QCD, as well as other asymptotically free QFTs at zero or finite density (like the Kondo problem).
There have been essentially very few investigations of such flows in the literature, \cite{gkn,hartnoll}, because they are not generic in top-down models, and they are explicitly avoided (because of their computational difficulty) in bottom up models. The only examples that have been studied do not seem to be exhaustive. They include, asymptotically-soft AdS minima of the potential, relevant for holographic models of YM, \cite{gkn}, that were argued recently to also be relevant in the cosmology of primordial inflation, \cite{cosmo}, and marginally relevant asymptotics of holographic Lifshitz fixed points, \cite{hartnoll}.

The purpose of this note is to analyze marginally relevant holographic flows in their full generality as they can arise in different contexts, under distinct realizations. They will also be compared to their field theory counterparts.

Moreover, the holographic approach allows the incorporation of non-perturbative corrections to the $\beta$-functions as well as the flow, due to non-trivial vacuum expectation values. We will calculate their form in full generality.

\bigskip
Our results are as follows:

\begin{enumerate}

\item We review the case of relevant operators, and the superpotential formalism in order to establish notations.

\item We show that the geometric singularities of a solution correspond to places where the potential $V$ and/or the superpotential $W$ diverge.

\item We define the holographic $\beta$-function, and derive a non-linear first-order equation for it, valid to leading order at strong coupling and large$-N$.

\item We analyze the case of classically marginal operators whose potential has zero mass term and the first non-trivial term is $\phi^n$, $n\geq 3$ in the regime $\phi\to 0$.
    The solutions around the UV fixed point ($\phi=0$) are derived and indicate that such operators have a perturbative $\beta$-function that starts at $n-2$ loop order.

\item We find the leading non-perturbative corrections, that behave exponentially in the inverse of the coupling constant.

\item We find however that the scalar diverges as it approaches the boundary, and this case seems inconsistent with the original assumptions ($\phi\ll 1$).

\item We also analyze classical marginal operators near a UV fixed point at $\phi\to \pm\infty$. This case corresponds to asymptotically-soft fixed points advocated for IHQCD,\cite{gkn}, and appears also in cosmology in Starobinsky-like models, \cite{staro,cosmo}. We find that they have always a non-trivial one-loop $\beta$-function, in close analogy to YM in four dimensions. Moreover, the scalar coupling asymptotes properly to the UV fixed point at the boundary of the geometry.

\item The leading non-perturbative corrections to the $\beta$-function are evaluated and shown to be similar to those expected in YM.

\item The multiscalar case is analyzed in a similar spirit. Here there is also a Zamolodchikov metric that enters the bulk description apart from the scalar potential. The perturbative RG flows are constructed near a UV fixed point, and it is shown that the contributions of the Zamolodchikov metric to the $\beta$-functions appear first at two-loops.

\end{enumerate}

We conclude  with an important open question to which we know the answer only in very special cases : which concrete cases of marginal operators arise from top-down string theory constructions at string tree level?

\section{Generic Perturbations Near the AdS Boundary}

We will start our study by setting up the general problem of RG flows of a scalar operator. To do this we will have to confine ourselves to Einstein-Dilaton (ED) theory which at the two derivative level\footnote{Generically this is justified at the strong coupling and large N limit, however adding higher derivatives of the scalar and the metric, does not affect the structure of the $\beta$ functions in the scaling region.}  after field redefinitions, takes the form:
\be
\label{1}
S_{2d} =M^{d-1}\int d^{d+1}x \sqrt{-g}
\biggl\{R -{1\over 2} \pa_{\mu}\phi\pa^{\mu}\phi+V(\phi)\biggl\}
\ee
Here $d$ is the space-time dimension of the QFT (living on the boundary), while the space-time dimension of the bulk ED gravity is $d+1$.

The metric as usual is dual to the stress tensor in the QFT and is always included, as it is always sourced even in the vacuum saddle-point (gravitational) solution\footnote{The basic rules of the holographic correspondence are detailed in several papers and textbooks, see eg. \cite{thebook}.}. We also have included a bulk scalar field $\phi$ dual to a scalar operator in the QFT.

In this section, there are no restrictions on the form of the potential, apart from the requirement of at least one extremum at $\phi=\phi_*$ which yields an $AdS_{d+1}$ geometry as the vacuum solution, with curvature scale $\ell$ and signals the existence of a CFT:
\be
V(\phi_*)={d(d-1)\over \ell^2} + \cdots
\label{2}
\ee
The general form of the equations of motion obtained from \eqref{1} is then:
\begin{align}
&
 \nabla^{\mu} \nabla_{\mu} \phi + V'(\phi) = 0, \label{eomphi}  \\
&\label{efe}
R_{\mu \nu} - {1 \over 2} g_{\mu \nu } \left( R - {1 \over 2 }  \left( \partial \phi \right) ^2 + V(\phi) \right) - {1 \over 2} \partial_\mu \phi \partial_\nu \phi  = 0.
\end{align}

\bigskip
\subsection{The Standard Perturbative Method}
\label{stdperturb}
\bigskip

We will proceed now to study RG flows in the neighborhood of the $AdS$ fixed point, (\ref{2}), also known in QFT as the scaling region.

In the \emph{conformal coordinate frame} for the metric,  the ansatz that we will be studying for the Poincar\'e-invariant field theory metric and the dilaton is:
\be
ds^2=e^{2A(r)}\left[dr^2+dx_{\m}dx^{\m}\right]\sp \phi(r)\,.
\label{3}
\ee
were the radial direction is the holographic direction and will be related to the RG scale as is standard in holography.

The $AdS_{d+1}$ solution:
\be
e^{2A}={\ell^2\over r^2}\sp \phi=0 \sp
\label{4}\ee
then corresponds to the (UV) CFT, with the boundary at $r=0$. Substituting the ansatz \eqref{3} into \eqref{eomphi} and \eqref{efe}, we obtain the equations for the scale factor and the scalar:
\begin{align}
&(d-1) \left[ A'(r)^2 - A''(r) \right] - {1 \over 2} \phi'(r)^2 = 0, \label{eq1} \\
&(d-1) A ''(r) + (d-1)^2 A'(r)^2 - e^{2A(r)} V(\phi) = 0, \label{eq2} \\
&e^{-2A(r)} \left[ \phi '' (r) + (d-1)A'(r) \phi'(r) \right] + {d V(\phi)  \over d \phi }= 0, \label{eq3}
\end{align}
where we have used the convention
$' \equiv {d \over dr}$.
These equations capture the solutions with Poincar\'e invariance in $d$ dimensions.

Equations (\ref{eq1})-(\ref{eq3}) can be solved perturbatively near the boundary. We choose the UV minimum to be at $\phi_*=0$ without loss of generality\footnote{It can be made to vanish by a shift in $\phi$ that does not affect the other terms in the action.}. We then expand the potential near the minimum to quadratic order\footnote{The convention chosen here for the mass term of the potential differs from \cite{aharony}. There it appears with a minus sign.}:
\be
V(\phi) = {d(d-1) \over \ell^2} +\half m^2 \phi^2+\cdots\sp \phi\to 0.
\label{pot}\ee
Working at next to leading order, we obtain:
\begin{align}
\label{fi}\phi(r) = & c_1 r^{\Delta_+} + c_2 r^{\Delta_-}+\cdots\, , \, \, \Delta_{\pm} = {d \pm \sqrt{d^2 - 4 m^2 \ell^2} \over 2}\, , \,\\
 A(r) =& \log {\ell \over r}  - {c_2^2 \Delta_- r^{2\Delta_-} \over 4(d-1)(1+ 2 \Delta_-)} - {c_1^2 \Delta_+ r^{2\Delta_+} \over 4(d-1)(1+ 2 \Delta_+)} - {c_1 c_2 \Delta_+ \Delta_- r^d \over d(d^2-1)}+\cdots
 \label{nexto}
\end{align}
with $c_{1,2}$ real.

We therefore derive the basic rule of the  $AdS/CFT$ dictionary, namely that
\be
m^2 \ell^2 = \Delta_+\Delta_-=\Delta \left( d - \Delta \right)
\ee
where $\Delta$ is the scaling dimension of the operator $\co$ dual to $\phi$ in the dual QFT.

There are several cases to be distinguished depending on the values of the mass:
\begin{enumerate}

\item ${d^2\over 4}>m^2\ell^2>0$ corresponds to relevant operators. If $m^2\ell^2>    {d^2\over 4} $,  the dimensions $\Delta_{\pm}$ are complex (and therefore this case is unacceptable). The scalar violates the BF bound.

    There are two important subcases:
\begin{itemize}

\item     $0<m^2\ell^2<{d^2\over 4}-1$ in which only $\Delta=\Delta_+$ and $c_1$ is the vev while $c_2$ is the source in (\ref{fi}).

\item     ${d^2\over 4}>m^2\ell^2>{d^2\over 4}-1$, where there are two quantizations of the scalar field as both linearly independent solutions are normalizable near the boundary. In one, $\Delta=\Delta_+$ and $c_{1,2}$ are as above, while in the other, $\Delta=\Delta_-$ and the role of $c_{1,2}$ is reversed.

\end{itemize}

\item $m^2<0$ corresponds to irrelevant operators.

\item $m=0$ corresponds to operators that to this order are marginal.

\end{enumerate}

\bigskip
\subsection{The Superpotential Method}
\label{supmethod}
\bigskip

There is however a more convenient formalism for our purposes, \cite{gkn,eff}, which uses a different set of coordinates, the \emph{domain wall coordinates}.
This is convenient because we will write the radial evolution equations for the scalar as first order RG equations in terms of a $\beta$-function.

  We use a new bulk coordinate $u$, related to the previous one through the relation $r = \ell e^{-u/\ell }$ in the $AdS$ case. Hence the boundary of an asymptotically $AdS$ solution now lies at $u = + \infty$. Using these coordinates, the ansatz (\ref{3}) for the metric takes the form:
\be
ds^2 = du^2 + e^{2A(u)} dx_{\mu} dx^{\mu}.
\ee
The $AdS_{d+1}$ solution is now:
\be
A(u) = { u \over \ell} \; , \; \phi(u) = 0.
\ee
Substituting the new form of the ansatz into the equations of motion \eqref{eomphi} and \eqref{efe}, we obtain:
\begin{align}
& d(d-1)\dot{A}(u)^2-{1\over 2}\dot\phi^2 - V(\phi) = 0, \label{eq11} \\
& 2(d-1) \ddot{A}(u) + \dot{\phi}(u)^2 = 0, \label{eq22} \\
& \ddot{\phi}(u) + d \dot{A}(u) \dot{\phi}(u) + { dV(\phi) \over d \phi}=0. \label{eq33}
\end{align}
We then introduce a function $W(\phi)$ that we  call the \emph{superpotential}\footnote{In supergravity actions it is indeed the superpotential, but it is a more general concept, and as shown in \cite{eff} it gives the boundary QFT effective potential for the scalar.}. It is a function of the scalar only but does not depend explicitly on the radial coordinate. The superpotential satisfies:
\be
\label{weq}
{dW^2\over 2(d-1)}-W'^2=2V(\phi).
\ee
where the prime is derivative with respect to $\phi$.
If (\ref{weq}) is valid then the first order equations:
\be
\label{1oeq}
\dot A = -{W\over 2(d-1)} \sp \dot\phi= {dW\over d\phi}\equiv W' \;,
\ee
are equivalent to the equations of motion (\ref{eq11}), (\ref{eq22}) and (\ref{eq33}).

The equations (\ref{1oeq}) are first order equations. Hence we only need one initial condition for each function $A$ and $\phi$. However, $W$ is determined through (\ref{weq}) which is also a a first order differential equation, and we therefore need one initial condition for $W$ to solve the equation uniquely.
 In total we have 3 initial conditions for this system of equations.

 This is the correct number of the initial conditions of the original set of equations (\ref{eq11})-(\ref{eq33}) as the $\phi$ equation is second order and has two arbitrary initial conditions while the equation for $A$ is first order, and has another integration constant.

The non-trivial integration constant of equation (\ref{weq}) controls  the (non-perturbative) expectation value of the operator dual to $\phi$.
Typically, as the equation (\ref{weq}) is non-linear, the space of solutions consists of two generic families that lead to singular solutions for the scalar, and a unique solution in between that gives a regular evolution for the scalar, \cite{thermo, hr}.

\subsection{Perturbative Calculation of the Superpotential}

The equation (\ref{weq}) can be solved perturbatively near the boundary ($\phi\to 0$) as a regular expansion in $\phi$:
\be
W(\phi) = W_0 + W_1 \phi +  W_2 \phi^2 +  \cdots.
\ee
Expanding to second order and using the potential (\ref{pot}), we obtain the following set of equations:
\be
{d W_0^2 \over 2 (d-1)} - W_1^2  = {2d(d-1) \over \ell^2} \sp {d W_1 W_0 \over (d-1) } - 4 W_1 W_2 = 0 \sp
\ee
\be
{dW_1^2\over 2(d-1)}+{dW_0W_2\over d-1}-4W_2^2-6W_1W_3=m^2.
\ee
which are solved in terms of $W_0$:
\be
W_1 = \pm  \sqrt{ {2d \over \ell^2 } - {dW_0^2 \over 2 (d-1)^2}} \sp W_2 = {dW_0 \over 4(d-1)}.
\label{e1}\ee
Hence, at quadratic order the expansion is:
\be
W(\phi) =  W_0 \pm  \sqrt{ {2d \over \ell^2 } - {dW_0^2 \over 2 (d-1)^2}} \phi +  {dW_0 \over 4(d-1)} \phi^2  +  O(\phi^3).
\label{ww}\ee
However, this solution is too general for our purposes and does not satisfy important holographic requirements. We need that after solving the first order equations using this superpotential, that we obtain the neighborhood of the boundary.
 We solve (\ref{1oeq}) using (\ref{ww}) to linear order,  $W(\phi) = W_0 + W_1 \phi$. The equation $\dot \phi = W_1$ gives:
\be
\phi(u) = W_1 (u - u_0).
\ee
where we have adjusted the boundary at $u=u_0$ so that $\phi$ vanishes there as it should.
For the metric, we are led to solve $\dot A = - W / 2(d-1)$ which implies
\be
ds^2=du^2+e^{A_0-{\left(W_1^2(u-u_0)+W_0\right)^2\over 2(d-1)W_1^2}}dx^{\mu}dx_{\m}.
\ee
It is clear, that $u=u_0$ is not a boundary for this metric (where the scale factor must diverge) unless $W_1=0$.
 Hence we must impose $W_1=0$ and the expansion (\ref{ww}) is no longer valid. Instead we  obtain
\be
W(\phi) = -{2 (d-1) \over \ell} -2{\Delta_{\pm}\over \ell}\phi^2+ O(\phi^{3}).
\label{e}\ee
Both overall signs for $W$ solve the equations. The minus sign corresponds to leaving a UV Fixed point. This can be seen from (\ref{1oeq}), and the fact that $u$ decreases away from the boundary at $u=+\infty$ and the fact that $A$ should decrease.

\subsection{The Holographic $\beta$-function}

In holography with a two-derivative bulk action, the scale factor $e^{A}$ plays the role of an RG scale. Apart from the fact that near the AdS boundary $r\to 0$ (or $u\to+\infty$) it coincides with the radial coordinate, it can be shown from the equations (\ref{eq11})-(\ref{eq33}) that it is a monotonic function, that diverges in the UV and vanishes in the IR, \cite{zafa}.
We may therefore associate the exponent of the scale factor $A\leftrightarrow \log \mu$ to the logarithm of the RG scale.

In view of this we may define the holographic analogue of the $\beta$ function as

\be
{d\phi\over dA}\equiv \beta_{\phi}=-{2(d-1) \over W} {d\over d\phi}W(\phi)\;,
\ee
where we have used the first order equations of motion (\ref{1oeq}) in the righthand side above.

The nonlinear equation for the superpotential (\ref{weq}) implies a non-linear equation for the $\beta$-function.
 Differentiating \eqref{weq} and using \eqref{1oeq}, we obtain a first order non-linear  differential equation for the holographic $\b$-function:
\be
\b = (d-1)  {d \over d \phi} \log \left( {\b^2 - 2d(d-1)  \over V }\right).
\label{nonl}\ee

\subsection{Regularity of the Solutions}

Taking $\phi$ as a coordinate instead of $u$, the metric becomes:
\be
ds^2={d\phi^2\over W'^2}+e^{2A}(dx_{\m}dx^{\m}),
\ee
where we have used the equations of motion \eqref{1oeq}. The Ricci scalar can be computed to be
\begin{align}
R =&{d\over 4(d-1)}\left[W^2-8V\right].
\label{r1}
\end{align}
where we have once again used the equations of motion and where all derivatives are with respect to $\phi$. On the other hand, taking the trace of the Einstein equations (\ref{efe}) yields:
\be
R = {1\over 2}(\pa\phi)^2-{d+1\over d-1}V.
\ee
Substituting this back into the Einstein equations allows us to obtain a simple expression for the Ricci tensor:
\be
R_{\m\n}={1\over 2}\pa_{\m}\phi\pa_{\n}\phi-{V\over d-1}g_{\m\n}.
\ee
From this we can readily deduce the following invariant:
\begin{align}
R_{\m\n}R^{\m\n} =&{d\over 16(d-1)^2}\left[16(d+1)V^2-8d VW^2+dW^4\right].  \label{r2}
\end{align}
Finally, we can calculate directly the Riemann invariant:
\begin{align}
R_{\m\n\r\s}R^{\m\n\r\s}= &{d\over 8(d-1)^3}\left[32(d-1)V^2-16(d-1)VW^2+(2d-1)W^4\right].
\label{r3}\end{align}

From these three formulae (\ref{r1}), (\ref{r2}) and (\ref{r3}) we can conclude the following concerning the regularity of the metric (finiteness of the curvature invariants):
\begin{enumerate}

\item If at a point $\phi$ both V and W are finite, then the geometry is regular.

\item If at a point $\phi$, $V$ diverges but $W$ remains bounded, or vice versa, this is a curvature singularity.

\item If $V\to \infty$ as $\phi\to \phi_*$, then in order for the scalar curvature to be regular , $W$ should also diverge as
$W^2\sim 8V$. However in that case the Ricci square is always divergent.

\end{enumerate}
We conclude that the geometric singularities correspond to places where $V$ and/or $W$ diverge.

If we now assume that we have an Einstein space, i.e. that $R_{\m\n}=\Lambda(\phi)g_{\m\n}$, then from (\ref{r1}) and (\ref{r2}) we obtain:
\begin{align}
& \Lambda^2(\phi) (d+1) = {d\over 16(d-1)^2}\left[16(d+1)V^2-8d VW^2+dW^4\right]\\
&  \Lambda(\phi) (d+1) = {d\over 4(d-1)}\left[W^2-8V\right].
\end{align}
Eliminating $W$ out from these equations gives:
\be
V(\phi)=-(d-1)\Lambda.
\ee
In particular in the case of AdS space $\Lambda$ is constant and then the potential must be constant.

\section{Marginal Operators: The Cubic and Higher Orders Cases}
\label{cubic}

In this section we will assume that  the extremal point of the potential is degenerate, $m=0$ and therefore the operator is marginal to leading order.
We will also assume that the first non-trivial derivative comes in at order $n\geq 3$.

We use the superpotential method of section \ref{supmethod} to study the gravitational solutions with the following potential:
\be
V(\phi) = {d(d-1) \over \ell^2} + {g_n \over \ell^2} \phi^n+\cdots ,
\ee
near $\phi\to 0$ where we have chosen the critical point of this potential to be.

\label{super}

The equation (\ref{weq}) can be solved perturbatively near the boundary by expanding $W$ around the $AdS$ solution $\phi=0$:
\be
W(\phi) = \sum_{n=0}^{\infty} W_n \phi^n .
\ee
We obtain\footnote{We assume here the presence only of the $\phi^n$ term in the potential in order to track which terms in the $\beta$-function are affected.}:
\be
  \label{wsol}
 W(\phi) = -{1\over \ell} \left( {2(d-1)} + {g_n \over d } \phi^{n} + {g_n^2 n^2 \over 2  d^3} \phi^{2n-2} +\d_{n,3} {54g_n^3 \over d^5}\phi^{2n-1} \right)+O(\phi^{2n}).
 \ee

We may now use the superpotential found,  to compute the (perturbative) $\beta$-function as
\be
 \beta_{\phi}\equiv {d\phi\over dA}=-{2(d-1) \over W} {d\over d\phi}W(\phi).
\ee
Using the perturbative solution (\ref{wsol}) for $W$, we find to next to leading order:
\be
 \beta_\phi = - {g_n n \over d} \phi^{n-1} - {g_n^2 n^2 (n-1) \over d^3}\phi^{2n-3} + O(\phi^{2n-1}).
\label{50}\ee
From (\ref{50}) we can interpret this $\beta$ function as follows.
 If $n=2$, at lowest order, we obtain a linear term in $\phi$, proportional to $d- \Delta$. This the standard classical contribution to the $\beta$-function of non-marginal operators.

Marginal operators have no quadratic term around the fixed point. If the first non-trivial term is cubic ($n=3$), then the leading $\beta$ function is quadratic and this is similar to the QFT case where  a one-loop contribution is the leading contribution to the $\beta$-function.

When $n>3$, then this is similar to the field theory case, where the first non-trivial contribution arises at $n-2$ loops.
We also observe, that a single bulk $\phi^n$ interaction, generates an infinite number of terms in the perturbative $\beta$ functions as seen in (\ref{50}).
For $n=3$,  all higher terms appear, while for higher $n$ there are gaps at low number of loops. This is explicitly seen in appendix \ref{wsoll} where the superpotential and the $\beta$ function are computed to higher orders.

The standard perturbative method can be used to obtain explicit expressions for the metric\footnote{Here $A$ denotes the logarithm of the scale factor.} and scalar field. For a dual operator $\mathcal{O}$ with $\langle \mathcal{O} \rangle =0$, we obtain:
\begin{equation}
\phi(A) =\phi_0- {g_n n\over d} A \phi_0^{n-1} - {g_n^2 n^2(n-1) \over d^2} A \left( {1 \over d} - \half A \right) \phi_0^{2n-3} + O(\phi_0^{2n-1}),
\label{a222}\end{equation}
where $\phi_0$ is the infinitesimal source for $\phi$. We can also express the scale factor in terms of the radial coordinate $r$:
\begin{equation}
A(r) = \log {\ell \over r} - {g_n \over 2 d(d-1)} \phi_0^n + {g_n^2 n^2 \over 4(d-1) d^3} \left[ 1- 2 d \left( 1- \log {r \over \ell} \right) \right] \phi_0^{2n-2} + O(\phi_0^{2n-1}) .
\end{equation}

\subsubsection*{Non-perturbative Corrections}
We will now try to extend this perturbative solution by including non-perturbative terms that originate in non-trivial vevs for the bulk scalar. In the case of a massive scalar field, it is known that the full non-perturbative solution is expressed as follows, \cite{eff} :
\be
W(\phi) = \sum_{k = 0}^{\infty} W_k(\phi) \sp W_k(\phi) = \mathcal C^k \phi^{k \d} \sum_{j = 0}^{\infty} A_{k,j} \phi^{j} \sp
\label{d1}\ee
where $\d$ is equal to $d/\Delta_-$, and where $\mathcal C$ is related to the vev $\langle \mathcal O \rangle$. $W_0(\phi)$ denotes the perturbative power series solution, valid in the $\phi\to 0$ regime.

For the marginal case, the non-perturbative contributions do not take the form of non-integer powers of $\phi$. To obtain the form of the possible non-perturbative terms, we will expand the superpotential again as in (\ref{d1}) and we will treat $W_{k>1}$ as small. This is equivalent to a small vev expansion, and we will formalise it by writing
\be
W = \sum_{k =0}^\infty x^k W_k.
\label{d2}\ee
and do perturbation theory in $x$ and then set it to one at the end.

Substituting (\ref{d2}) into \eqref{weq}, and separating powers of $x$, we obtain for any integer $k\geq 0$:
\be
{d \over 2 (d-1)} \sum_{i + j = k} W_i W_j - \sum_{i+ j = k } W_i' W_j' =  \d_{k,0} 2V.
\ee

At each order, we obtain a first-order linear equation for $W_k$ of the form:
\be
{d \over (d-1)}W_0 W_k - 2W_0' W_k' = f_k(\phi),
\ee
where $f_k(\phi)$ is a function depending on the $W_m$ and $W_m'$ for $0 \leq m \leq k-1$. The general solution to this equation is:
\be
W_k = \exp\left[-{d}\int {d\phi\over \beta_0(\phi)}\right] \left( \mathcal C _k - \int { f_k(\phi) \over 2 W_0'} \exp\left[{d}\int {d\tilde \phi \over \beta_0(\tilde \phi)}\right] d\phi \right).
 \label{nonpsol}
\ee
For the leading correction $W_1$, we obtain \eqref{weq}:
\be
{d \over d-1}W_0 W_1 - 2 W_0' W_1' = 0.
\label{nonpert1}
\ee
and therefore
\be
W_1 = \mathcal C _1\exp\left[-{d}\int_{0}^{\phi} {d\chi\over \beta_0(\chi)}\right],
\ee
where we have used the definition of the perturbative holographic $\b$-function:
\be
\b_0 = -2 (d-1) {W_0' \over W_0}.
\ee
The constant $\mathcal C _1$ is non-perturbative and can be fixed only if a global and regular flow solution for the scalar and the metric has been found. Similarly, for $k=2$, we obtain:
\be
{d \over (d-1)} W_0 W_2 - 2W_0' W_2' = W_1'^2 - {d \over 2(d-1)} W_1^2 \equiv f_2.
\ee
with solution
\begin{align}
W_2 =&\exp\left[-{d}\int_{0}^{\phi} {d\phi\over \beta_0(\phi)}\right]   \left( \mathcal C _2 + \int_{0}^{\phi} \left( {d \over \b_0^2} - {1 \over 2(d-1)} \right) {d(d-1) \over  \b_0} \mathcal C _1^2e^{ - \int_{0}^{\phi} \b_0 \left( {d \over \b_0^2} - {1 \over 2(d-1)}\right) }d \chi  \right)
\end{align}
where we have used:
\be
W_0=e^{-{1\over 2(d-1)}\int d\phi \beta_0(\phi)}\sp  W_1=\mathcal C _1 e^{-d\int {d\phi\over \beta_0}} \sp W_0' = - {\b_0 \over 2(d-1)} e^{- {1 \over 2(d-1)} \int d\phi \b_0}.
\ee
Note that the integration constant $\mathcal C _2$ can be absorbed into $\mathcal C _1$ in $W_1$. This is as it should, as the full solution should contain a single undetermined constant as the equation for the superpotential is of the first order.

By induction it is straightforward  to see that all the higher terms $W_i$, $i \geq 3$ can also be expressed in terms of integrals of $\b_0$.
We will now compute explicitly $W_1$, using the perturbative form \eqref{wsol} obtained for $W_0$. From \eqref{nonpert1}, we have:
\be
{d \log W_1 \over d \phi } = {d \over 2(d-1)} {W_0 \over W_0'} = {d^2 \over g_n n \phi^{n-1}} - {n-1 \over \phi} - \d_{n,3} {18 g_n \over d^2} + O(\phi),
\ee
which is solved by:
\begin{align}
W_1(\phi) &= \mathcal C _1 e^{- {d^2 \over g_n n(n-2) \phi^{n-2} }-(n-1) \log \phi -  \d_{n,3}{18g_n\over d^2}\phi+\cdots }.
\label{W1+}
\end{align}

It should be noted here that the non-perturbative solution is captured completely in the non-linear equation for the $\beta$ function (\ref{nonl}) we derived earlier. Also the arbitrary integration constant that appears in the non-perturbative part, is eventually fixed by demanding that the superpotential gives rise to a globally regular solution. This constraints the integration constant to take a restricted set of values, typically only one, \cite{eff}.

We can also compute the contribution $\b_1$ of the non-perturbative part of the superpotential to the $\b$-function:
\begin{align}
\b =& -2(d-1) {d \over d \phi} \log W = \underbrace{-2(d-1) {d \over d \phi} \log W_0}_{\b_0} \underbrace{- 2(d-1) {d \over d \phi} {W_1 \over W_0}}_{\b_1}+ \cdots
\end{align}
with
\be
\beta_1=
-2(d-1)\left({W_1'\over W_0}-{W_0'\over W_0}{W_1\over W_0}\right)=\left({2d(d-1)\over \beta_0}-\beta_0\right){\mathcal C_1\over W_0}e^{-d\int {d\phi\over \beta_0}}.
\ee

We obtain:
\begin{align}
\b_1 =& \ell \mathcal C _1 e^{- {d^2 \over g_n n (n-2) \phi^{n-2}} - (n-1) \log \phi - \d_{n,3} {18 g_n \over d^2}\phi} \nn \\
& \times \Big( {d^2 \over g_n n \phi^{n-1}}  -{n-1 \over \phi} - \d_{n,3} {18 g_n \over d^2} + O(\phi) \Big).
\end{align}

Therefore for $n=3$ corresponding to a one-loop $\beta$-function we obtain
\be
\beta_1\simeq \mathcal C {d^2\over 3g_3\phi^2}e^{- {d^2 \over 3g_3 \phi}}.
\ee
which has the typical non-perturbative form we are familiar with from QCD.

\subsubsection*{Asymptotic Behavior of the Coupling }

We can now use the expression for $\beta_\phi$ to discuss the asymptotic behaviors depending on the coupling constant $g_n$. For this, we integrate (\ref{50}) in the following way:
\begin{align}
& \int_{\phi(u_0)}^{\phi(u)} {d \phi \over \beta_\phi} = \int_{A(u_0)}^{A(u)} dA
\end{align}
where $u_0$ is some arbitrary scale, and which then gives:
\begin{align}
&  - {d \over g_n n}\int_{\phi(u_0)}^{\phi(u)} {1 \over \phi^{n-1}} \left( 1 - {g_n n (n-1) \over d} \phi^{n-2} + O(\phi^n) \right) d\phi = A(u) - A(u_0).
\end{align}
For $n>2$, the result is:
\begin{align}
{d \over g_n n (n-2)} \left( {1 \over \phi^{n-2}(u) } - {1 \over \phi^{n-2}(u_0) } \right) + (n-1) \log \left( {\phi(u) \over \phi(u_0)} \right) + O(\phi^2)= A(u) - A(u_0).
\end{align}
Since the UV (or near-boundary limit) limit is at $u \rightarrow + \infty$ and $A(u) \sim  u/\ell$, we observe that the theory is asymptotically free for $g_n >0$. The IR limit is at $u \rightarrow - \infty$ and this time the theory will be IR free for $g_n < 0$.

Inverting the above formula, one obtains:
\be
\phi
\simeq  \phi_0 - {g_n n \over d} \phi_0^{n-1} A  + {g_n^2 n^2 (n-1) \phi_0^{2n-3} \over 2 d^2}  (A^2+{\cal O}(A)) + {\cal O}(\phi_0^{2n-2}),
\label{a223}\ee
in agreement with (\ref{a222}).

We summarize the solution as follows: we have set our UV fixed point to $\phi=0$ without loss of generality, and expanded the potential near this fixed point, with $\phi\ll 1$. We have found that near the boundary, $A\to\infty$,  the solution for $\phi$ is given by (\ref{a222}) or (\ref{a223}) and therefore, $\phi\to\infty$ as we approach the boundary, $A\to \infty$. This contradicts our assumptions and therefore this case is inconsistent.

\subsubsection*{Single-term $\b$-function}

We have seen that a single non-trivial term in the bulk scalar potential produces an infinite number of terms in the $\beta$ function.
In this part we will ask the opposite question: are there bulk potentials that give a (perturbative)  $\beta$ function containing as single term?

This is equivalent to imposing:
\be
- 2(d-1) {W' \over W} = - k \phi^{n-1}
\ee
which can be solved by :
\be
W(\phi) = \mathcal{C} e^{{ k \over 2n(d-1)}\phi^n},
\ee
where $\mathcal{C}$ is an integration constant. Substituting this result back into (\ref{weq}), we get:
\begin{align}
V(\phi) &= \half \left( {d \over 2(d-1)} \left( \mathcal{C} e^{{ k \over 2n(d-1)}\phi^n}\right)^2 - \left(\mathcal{C} { k \over 2(d-1)}\phi^{n-1} e^{{ k \over 2n(d-1)}\phi^n}  \right)^2 \right) \\
& = \half  \mathcal{C}^2 e^{{ k \over n(d-1)}\phi^n}   \left( {d \over 2 (d-1)} -{k^2 \over 4(d-1)^2} \phi^{2n-2} \right)
\end{align}
The potential therefore has Liouville-type interactions.

\bigskip
\section{Marginal Operators: Asymptotically Soft Fixed Points}
\bigskip

In our previous discussion we have assumed that the UV fixed point is a finite value of the scalar field. This is not the only option, and in this section we will describe the only alternative, namely asymptotically soft nearly marginal fixed points. This is the case utilized in holographic models of YM like Improved Holographic QCD (IHQCD), \cite{gkn,rev}, and is also relevant in cosmology (see \cite{cosmo}).

The critical point now will be at $\phi\to\pm \infty$. Without loss of generality we consider  the following potential:
\be
V(\phi)={d(d-1)\over \ell^2}\left[1+\sum_{n=1}^{\infty} V_n~\lambda^n\right]\sp \l\equiv e^{-a\phi}\sp  a>0 \sp
\ee
and we will be working in the region $\phi\to \infty$, so that $\lambda\to 0$.
$\l=0$ is an AdS fixed point. This potential has the property that around the AdS fixed point, all derivatives of the potential vanish justifying the term asymptotically soft fixed-point.

Solving perturbatively \eqref{weq}, we obtain for the first few coefficients:
\be
W=\sum_{n=0}^{\infty}W_n \lambda^n\sp W_0 =- {2(d-1) \over \ell}\sp  W_1 = W_0 {V_1 \over 2 } \sp
\ee
\be
W_2 = W_0\left({V_2 \over 2}  - \left[1 -   { 2  (d-1)a^2 \over d} \right]  {V_1^2  \over 8}\right),
\ee
where once again the sign of $W_0$ has been chosen so that the solution leads to an $AdS$ boundary.

The $\beta$-function is calculated as follows:
\be
\b_\l \equiv  {d \l \over dA} = -a \l {d\phi \over dA} = 2(d-1) a \l {W' \over W}.
\ee
Substituting the previous results found for the superpotential, we obtain:
\begin{align}
\b_\l =& - (d-1) a^2V_1\l^2 - 2(d-1) a^2  \left[ V_2  - \left[1 -   { 2  (d-1)a^2 \over d} \right]  {V_1^2  \over 2}  \right] \l^3 + \dots
\end{align}
This is again a $\beta$-function corresponding to a marginal coupling, with the difference that here there is no option on what is the leading term in the $\beta$ function. It always is the ``one-loop" term.

From here, we can compute the corrections to the metric and to the dilaton by solving \eqref{1oeq}. For the dilaton we obtain
\be
\dot \phi = - {1 \over a \l  }{d\l \over d u  } = -a W_1 \l -2a W_2 \l^2 + \dots,
\ee
which is solved by:
\begin{align}
\l(u) = {\ell \over a^2 (d-1) V_1 u} + { \ell^2 \over a^4  (d-1)^2 V_1  } \left(   \left[1 -   { 2  (d-1)a^2 \over d} \right]  {1  \over 2} - {2V_2 \over  V_1^2}\right){\log u \over u^2}.
\end{align}

We can then turn our attention to the metric, for which the equation to solve is:
\begin{align}
\dot A =& -{1 \over 2(d-1)} \left( W_0 + W_1 \l + W_2\l^2 + \dots \right) \nn \\
=&  {1 \over \ell} + {V_1 \over 2 \ell}  \l + {1 \over \ell }\left({V_2 \over 2}  - \left[1 -   { 2  (d-1)a^2 \over d} \right]  {V_1^2  \over 8}\right)\l^2 + \dots
\end{align}
This is solved by:
\begin{align}
A(u) =&  {u \over \ell } + {1 \over a^2 2  (d-1) V_1 } \log u \nn \\
& - { \ell \over a^4  2(d-1)^2   } \left(   \left[1 -   { 2  (d-1)a^2 \over d} \right]  {1  \over 2} - {2V_2 \over  V_1^2}\right){\log u \over u } \nn \\
&-  {\ell \over a^4 (d-1)^2} \left( -  {V_2 \over 2 V_1^2} + \left[ 1 - {2(d-1) a^2 \over d} \right]{1 \over 8}    \right) {1 \over u }.
\end{align}

Working again at next to leading order, we can also express $\lambda$ as a function of $A$:
\begin{align}
\lambda =  {1 \over a^2 (d-1) V_1  A} +  {1 \over a^4 (d-1)^2 V_1} \left( \left[ 1 - {2(d-1) a^2 \over d}\right] \half - {1 \over 2 V_1}- {2V_2 \over V_1^2} \right)  {\log  A \over A ^2 } \nn \\
\end{align}

Note that in this case, and unlike that of the previous section, the expansion of the potential, and the RG flow are compatible. $\lambda$ remains small during the flow.

Using the setup in section \ref{super}, where $W_{np}$ denotes the leading non-perturbative correction to the superpotential, we are led to consider the following equation:
\be
W_0 W_{np} - {2(d-1) \over d} W_0' W_{np}' = 0,
\ee
We obtain
\begin{align}
{d \log W_{np} \over d \phi} = - {d \over (d-1) a V_1 \l}  - {d \over (d-1) a} \left( 1 - {2 V_2 \over V_1^2 } - {(d-1) a^2 \over d } \right)+\cdots,
\end{align}
which is integrated to
\be
W_{np}(\l) = \mathcal C~ e^{- {d \over (d-1) a^2 V_1 \l} }~~ \l^{{d \over (d-1) a^2} \left( 1 - {2 V_2 \over V_1^2 } - {(d-1) a^2 \over d } \right)}+\cdots.
\label{nonpertasy}
\ee

$W_{np}$ affects the expression for the scalar field. The equation to solve is:
\begin{align}
-{\dot \l \over a \l} = -a \left( W_{1} \l + 2 W_{2} \l^2 \right) - {d \over (d-1) a  V_1 \l}  W_{np}+\cdots,
\end{align}
where we have only kept the leading non-perturbative term in \eqref{nonpertasy}. This gives:
\begin{align}
\l(u) =& {\ell \over a^2 (d-1) V_1 u }  + { \ell^2 \over a^4  (d-1)^2 V_1  } \left(   \left[1 -   { 2  (d-1)a^2 \over d} \right]  {1  \over 2} - {2V_2 \over  V_1^2}\right){\log u \over u^2} \nn \\
& - {\ell \over (d-1)} \left( 1 + {\ell \over  d V_1 ^2 u} + {2 \ell^2  \over d^2  u^2 }  \right) \mathcal C _1 e^{-{d u \over \ell }}+\cdots.
\end{align}

Finally, we compute the contribution of the non-perturbative term to the $\b$-function. We have this time:
\begin{align}
\b_\l \equiv & {d \l \over dA } = - a \; {d\phi \over d A} = 2(d-1) a \l {d \over d \phi} \log W \nn \\
=& 2(d-1) a \l {d \over d \phi} \log \left( W_{pert} + W_{np} \right) =  2(d-1) a \l {d \over d \phi} \left( \log W_{pert}  + \log \left[ 1 + {W_{np} \over W_{pert}}\right] \right)\;.
\end{align}
We are interested in computing the second term, namely:
\be
\b_{\l,np} \equiv   2(d-1) a \l {d \over d \phi} {W_1 \over W_0 }
=  {\ell d \l  \over (d-1)} W_{np} \left[  {1 \over V_1 \l} +  \half - {2V_2 \over V_1^2 } - {(d-1)a^2 \over d}  \right]+\cdots.
\ee
$$
=\mathcal C~ {\ell d \l  \over (d-1)} ~ \left[  {1 \over V_1 \l} +  \half - {2V_2 \over V_1^2 } - {(d-1)a^2 \over d}  \right]~ e^{- {d \over (d-1) a^2 V_1 \l} }~~ \l^{{d \over (d-1) a^2} \left( 1 - {2 V_2 \over V_1^2 } - {(d-1) a^2 \over d } \right)}+\cdots
$$

This is in agreement with general dimensional expectations.

We summarize, that this second case of marginal operators seems self-consistent and described flows similar to those of asymptotically-free gauge theories in the weakly coupled regime.

Its realization in string theory so far can be found in higher derivative contexts, several possibilities of which have been analyzed in \cite{dis}, (see also \cite{ke} for a discussion of supergravity embeddings).

\section{The Multiscalar Case}

In the generic case, a single scalar operator, even if it is the only one sourced in the UV, mixes as it runs with other scalar operators. This is captured by a multi-scalar general action that we will investigate in this section.

The important difference is that apart from a bulk potential, there is a scalar-field metric now that also controls the holographic RG flow.
One of our goals will be to see at what order in the scaling region the metric affects the flow.

We consider therefore the case of $N$ scalars. We start from a modified version of \eqref{1}:
\be
S_{2d} = M^{d-1} \int d^{d+1}x \sqrt{-g} \left\{ R - \half G_{ij}(\phi) \p_\m \phi^i \p^\m \phi^j + V(\phi) \right\}.
\ee
The bulk equations can be easily deduced from previous calculations:
\be
R_{\m\n} - \half g_{\m\n} \left( R - \half G_{ij}(\phi) \p_\m \phi^i \p^\m \phi^j + V(\phi) \right) - \half G_{ij}(\phi) \p_\m \phi^i \p^\m \phi^j = 0 \sp
\ee
\be
\nabla^\mu \left( G_{ij} \nabla_\m \phi^j \right) + {\p V \over \p \phi^i} - \half {\p G_{kl} \over \p \phi^i} \p_\m \phi^k \p^\mu \phi^l = 0.
\ee

From this, using the ansatz in the domain wall frame, we obtain:
\be
d(d-1) \dot A ^2 + (d-1) \ddot A - V = 0 \sp  \label{eomM1}
\ee
\be
 2(d-1) \ddot A + G_{ij} \dot \phi ^i \dot \phi^ j = 0 \sp
\label{eomM2}
\ee
\be
 {\Gamma'}^i_{jk} \dot \phi^j \dot \phi^k + \ddot \phi^i + d \dot A  \dot \phi ^i + G^{ij}{\p V \over \p \phi^j} = 0 \sp
 \label{EinsteinM}
\ee
 where $\Gamma'$ are the  Christoffel symbols for the field metric.

A generalization of the first order equations \eqref{weq} and \eqref{1oeq} to the case of multiple scalars is:
\begin{align}
& \dot A = - {W \over 2 (d-1)} \sp \dot \phi ^i = G^{ij} {d W \over d \phi^j} \sp \label{1oeqM}\\
&{d W^2 \over 2  (d-1)} - G^{kl} {d W \over d \phi^k} {d W \over d \phi ^l } = 2 V. \label{weqM}
\end{align}
These reproduce the equations of motion \eqref{eomM1}, \eqref{eomM2} and \eqref{EinsteinM}.

In order to compute the superpotential and the corresponding $\b$-function, we will choose Riemann normal coordinates (RNC) around an $AdS$ extremum. With this choice, the potential takes the following form:
\be
V(\phi) = {d(d-1) \over \ell^2}  + \half{M^2}_{ij} \phi^i \phi^j +g_{ijk}\phi^i\phi^j\phi^k\cdots \;.
\label{normalp}
\ee Using the properties of the RNC, we can expand the metric around the $AdS$ extremum and obtain:
\be
G_{ij} = \d_{ij} + \half \phi^k \phi^l  \left. {d^2 G_{ij} \over d \phi^k d \phi^l} \right|_{\phi = 0}.
\label{metricexpand}
\ee
Expressing the second derivatives of the metric in terms of the Riemann tensor, we obtain:
\be
G^{ij} = \d^{ij} - {1\over 3} \d^{im} \d^{jn} R_{ mk  n l}\phi^{k}\phi^l + O(\phi^3).
\label{normalinverse}
\ee

Using the set up of the previous section, we can solve perturbatively \eqref{weqM}. The superpotential $W$ takes the general form:
\be
W(\phi) = W_0 + \sum_{i = 1}^N W_i \phi^i + \sum_{i,j=1}^N W_{ij}\phi^i \phi^j + \sum_{i,j,k=1}^N W_{ijk} \phi^i \phi^j \phi^k + O(\phi^4),
\ee
where the convention is that:
\be
W_{i_1 \dots i_k  } = \left. {d^k W \over d\phi^{i_1} \cdots d\phi^{i_k}} \right|_{\phi=0}.
\ee

As in the single scalar case, we should put the first derivatives to zero, in order for the vanishing of the scalars to happen at the extremum.
Solving perturbatively the equation \eqref{weqM} for the superpotential, we obtain the following set of equations:
\begin{align}
&{d \over 2(d-1)} W_0^2  = {2 d(d-1) \over \ell^2} \\
& {d \over (d-1)} W_0 W_{ij}  - \d^{kl} 4W_{jk}W_{il} = M_{ij}^2 \\
& {d \over (d-1)} W_0\sum_{i,j,p=1}^N W_{ijp} \phi^i \phi^j \phi^p - 12  \d^{kl}  \sum_{p=1}^N W_{pk} \phi^p  \sum_{i,j=1}^N W_{ijl} \phi^i \phi^j = 2 g_{i,j,k}\phi^{i}\phi^{j}\phi^{k}
\end{align}
which can be simplified to:
\begin{align}
& W_0 = - {2(d-1) \over \ell} \label{M1}\\
& -{2d  W_{ij} \over \ell}  - \d^{kl} 4W_{il}W_{kj} = M_{ij}^2 \label{M2} \\
& -{2d \over \ell} W_{ijp} - 12 \d^{kl} W_{k(p} W_{ij)l} = 2\d_{n,3} g_{ijp} \label{M3}
\end{align}
where indices are raised and lowered using the flat Euclidean metric since all quantities are evaluated at the center of the RNC, and where the sign of $W_0$ is chosen in order to recover an $AdS$ solution when all the scalar fields vanish.

Without loss of generality, we can take $M^2_{ij}$ and $W_{ij}$ to be real and symmetric. It is then straightforward to show using \eqref{M2} that $\left[ M^2, W \right] = 0$. Hence there exists a coordinate system in the scalar fields space in which both these matrices can be simultaneously diagonalized by a constant $O(N)$ rotation. With this choice, we can write:
\be
M^2 = \text{diag} \left( m^2_1, m^2_2, \dots, m^2_N \right).
\ee
The eigenvalues are real, and some of them can be equal to zero, in which case this would correspond to a marginal operator. Denoting $w_i$ the eigenvalues of $W_{ij}$, the equation  $\eqref{M2}$ can be written as:
\be
 4w_i^2 + {2d \over \ell}w_i +  m^2_i=0,
\ee
which is solved by:
\be
w_i = {- d\pm \sqrt{d^2 - 4 \ell^2 m_i^2} \over 4\ell }.
\ee

We now can define the holographic version of the $\b$-function for the case of multiple scalars:
\be
\b_\phi ^i \equiv  {d \phi^i \over d A} = - 2(d-1) G^{ij}  {d \over d \phi^j} \log W.
\ee
Using \eqref{normalinverse}, we obtain:
\begin{align}
\b^i_\phi =&2 \ell w_i \phi^i + 3\ell  \sum_{p,q=1}^N W_{ipq} \phi^p \phi^q \nn \\
&+ \sum_{p,q,r=1}^N \left( 4 \ell  W_{ipqr} + { \ell^2 \over (d-1)}  w_pw_i\delta^{pq}\delta^{ir} - {2 \ell \over 3} w_rR_{iprq} \right) \phi^p \phi^q \phi^r + O(\phi^4).
\end{align}
From this last equation and \eqref{M1}, \eqref{M2} and \eqref{M3}, we conclude the following

\begin{itemize}

\item  If a subset of operators are marginal, then their leading $\beta$-function starts generically at one-loop.

\item The mixing with relevant operators is also happening first at the one-loop. This persists even when all operators are marginal.

\item The field space metric (Zamolodchikov metric), equivalent to the curvature of the field space first affects the $\beta$-functions at two-loop level.

\end{itemize}

\section{Acknowledgements}\label{ACKNOWL}

We would like to thank J. Gauntlet, N. Warner and K. Pilch  for correspondence.

This work was supported in part by European Union's Seventh Framework Programme under grant agreements (FP7-REGPOT-2012-2013-1) no 316165,
PIF-GA-2011-300984, the EU program ``Thales'' MIS 375734, by the European Commission under the ERC Advanced Grant BSMOXFORD 228169and was also co-financed by the European Union (European Social Fund, ESF) and Greek national funds through the Operational Program ``Education and Lifelong Learning'' of the National Strategic Reference Framework (NSRF) under ``Funding of proposals that have received a positive evaluation in the 3rd and 4th Call of ERC Grant Schemes''. The research leading to these results has also received funding from the People Programme  (Marie Curie Actions) of the European Union's Seventh Framework Programme FP7/2007-2013/ under REA Grant Agreement No 317089 (GATIS).

\newpage
\appendix
\renewcommand{\theequation}{\thesection.\arabic{equation}}
\addcontentsline{toc}{section}{Appendix\label{app}}
\section*{Appendices}

\section{Perturbative Calculations for $n= 3,4,5$. \label{wsoll}}
\subsection{$n=3$}
\begin{align}
W(\phi) =& - {2(d-1) \over \ell}- { g_3 \over d \ell} \phi^3 - {9 g_3^2 \over 2 d^3 \ell} \phi^4 - {54 g_3^3 \over d^5 \ell} \phi^5 + g_3^2{ d^5 + 3888  (1-d) g_3^2 \over 4(d-1) d^7 \ell} \phi^6 \nn \\
& + 27 g_3^3 {d^5 + 3312 (1-d)g_3^2 \over 4 (d-1) d^9 \ell}  \phi^7 + 27 g_3^4 {119 d^5 + 362016  (1 -d)g_3^2 \over 16 (d-1)d^{11} \ell}\phi^8  \nn \\
&- {g_3^3 \over 8 (d-1)^2 d^{13} \ell} \left(d^{10} + 53460 d^5  (1-d)g_3^2 - 304 850 304 d g_3^4 + 152425152 (1+ d^2)g_3^4  \right)\phi^9 \nn \\
& - {27 g_3^4 \over 16 (d-1)^2 d^{15} \ell} \left( 5 d^{10} + 145152 d^5  (1-d)g_3^2 + 392097024  (1+d^2)g_3^4 - 784194048 d g_3^4 \right) \phi^{10}
\end{align}
\begin{align}
\beta_\phi =& - {3 g_3 \over d} \phi^2 - {18 g_3^2 \over d^3} \phi^3 - {270 g_3^3 \over d^5}\phi^4 + 3 g_3^2{ d^5 + 1944 (1 -d)g_3^2 \over (d-1) d^7}\phi^5 + 63 g_3^3 {d^5 + 2484  (1-d)g_3^2 \over (d-1) d^9} \phi^6 \nn \\
&+ 81{23 d^5 g_3^4 + 60336  (1-d)g_3^6\over (d-1) d^{11}} \phi^7 \nn \\
& - 3 g_3^3 {d^{10} + 21870 d^5  (1-d)g_3^2 + 57159 432 (1+d^2)g_3^4 - 114 318 864 d g_3^4 \over (d-1)^2 d^{13}}\phi^8 \nn \\
&- 135 g_3^4 {d^{10} + 19188 d^5  (1-d)g_3^2 + 49012128 (1+d^2)g_3^4 - 98024256 d g_3^4 \over (d-1)^2 d^{15}} \phi^9
\end{align}

\subsection{$n=4$}
\begin{align}
W(\phi) =& - {2(d-1) \over \ell} - {g_4 \over d \ell} \phi^4 - {8 g_4^2 \over d^3 \ell} \phi^6 + g_4^2{d^3 +768  (1 -d)g_4  \over 4 (d-1) d^5 \ell} \phi^8 + 12 g_4^3 {d^3 + 608  (1-d)g_4 \over (d-1) d^7 \ell} \phi^{10} \nn \\
& - g_4^3 {d^6 + 5504 d^3  (1-d)g_4 + 2924544 (1+d^2) g_4^2 - 5849088 d g_4^2 \over 8(d-1)^2 d^9 \ell} \phi^{12} \nn \\
&-3 g_4^4 {5d^6 + 15424 d^3(1-d)g_4 + 7409664 (1+d^2) g_4^2 - 14819328 d g_4^2\over (d-1)^2 d^{11} \ell} \phi^{14} \nn \\
& + {g_4^4 \over 64 (d-1)^3 d^{13} \ell } \left[ 5d^9 + 100352 d^6  (1-d)g_4 -452984832 d^4 g_4^2 \right. \nn \\
&+ 226 492 416 d^3(1+d^2) g_4^2 + 100317265920 (1-d^3) g_4^3 \nn \\
& - \left. 300 951 797 760 d(1-d) g_4^3  \right] \phi^{16}
\end{align}
\begin{align}
\beta_\phi =& - {4 g_4 \over d} \phi^3 - {48 g_4^2 \over d^3} \phi^5 + 4 g_4^2 {d^3 + 384 (1-d)g_4 \over d^5 (d-1)} \phi^7 + 160 g_4^3 {d^3 + 456  (1-d)g_4 \over (d-1) d^7} \phi^9 \nn \\
&-4 g_4^3 {d^6 + 2400 d^3 (1-d)g_4 + 1096704 (1+d^2)g_4^2 - 2193408 d g_4^2 \over (d-1)^2 d^9} \phi^{11} \nn \\
&- 336 g_4^4 {d^6 + 2112 d^3  (1-d)g_4 + 926208  (1+d^2)g_4^2 - 1852416 d g_4^2 \over (d-1)^2 d^{11}}\phi^{13} \nn \\
& +{4g_4^4 \over (d-1)^3 d^{13}} \left[ d^9 + 8192 d^6  (1-d)g_4 - 30081024 d^4 g_4^2 + 15 040 512 d^3(1+d^2) g_4^2 \right. \nn \\
& \left. + 6269 829 120 (1-d^3)g_4^3 - 18 809 487 360 d g_4^3 (1-d) \right]  \phi^{15}
\end{align}

\subsection{$n=5$}
\begin{align}
W(\phi) =& - {2(d-1) \over \ell} - {g_5 \over d \ell} \phi^5 - {25 g_5^2 \over 2 d^3 \ell} \phi^8 + {g_5^2 \over 4 (d-1) d^2 \ell} \phi^{10} \nn \\
& - {500 g_5^3 \over d^5 \ell} \phi^{11} + {75 g_5^3 \over 4 (d-1) d^4 \ell} \phi^{13} - {32500 g_5^4 \over d^7 \ell} \phi^{14} - {g_5^3 \over 8(d-1)^2 d^3 \ell} \phi^{15} \nn \\
& + {28 125 g_5^4 \over 16(d-1) d^6 \ell}\phi^{16} - {2825 000 g_5^5 \over d^9 \ell } \phi^{17} - {375 g_5^4 \over 16 (d-1)^2 d^5 \ell } \phi^{18} + {198125 g_5^5 \over (d-1)d^8 \ell} \phi^{19} \nn \\
& + 5 g_5^4 {d^7 + 3849 600 000  (1-d^3)g_5^2 - 11 548 800 000 d  (1-d)g_5^2 \over 64 (d-1)^3 d^{11} \ell} \phi^{20}
\end{align}
\begin{align}
\beta_\phi =& - {5 g_5 \over d} \phi^4 - {100 g_5^2 \over d^3} \phi^7 + {5 g_5^2 \over (d-1) d^2}\phi^9 - {5500 g_5^3 \over d^5} \phi^{10} + {325 g_5^3 \over (d-1) d^4 }\phi^{12} \nn \\
& - {455000 g_5^4 \over d^7} \phi^{13} - {5 g_5^3 \over (d-1)^2 d^3} \phi^{14} + {32750 g_5^4 \over (d-1) d^6} \phi^{15} - {48 025 000 g_5^5 \over d^9}\phi^{16} \nn \\
&- {675 g_5^4 \over (d-1)^2 d^5} \phi^{17} + {4132500 g_5^5 \over (d-1) d^8} \phi^{18} \nn \\
&+ 5g_5^4 {d^7 + 1203 000 000  (1-d^3)g_5^2 - 3609 000 000 d  (1-d)g_5^2 \over (d-1)^3 d^{11}} \phi^{19}
\end{align}

\end{document}